\documentclass[aps,prl,twocolumn,showpacs]{revtex4}
\usepackage{amssymb}
\usepackage{graphicx,amsmath}

\newcommand{\bea}{\begin{eqnarray}}
\newcommand{\eea}{\end{eqnarray}}
\newcommand{\ben}{\begin{equation}}
\newcommand{\een}{\end{equation}}
\newcommand{\benu}{\begin{enumerate}}
\newcommand{\enu}{\end{enumerate}}

\newcommand{\al}{\alpha}

\newcommand{\om}{\omega}
\newcommand{\Om}{\Omega}
\newcommand{\ep}{\epsilon}

\newcommand{\si}{\sigma}
\newcommand{\dl}{\delta}

\newcommand{\ham}{\mathcal{H}}
\newcommand{\ord}{\mathcal{O}}

\newcommand{\cda}{c^{\dagger}}
\newcommand{\fda}{f^{\dagger}}
\newcommand{\Fda}{F^{\dagger}}
\newcommand{\bda}{b^{\dagger}}
\newcommand{\bk}{{\bf k}}

\begin{document}
\title{Signature of Kondo breakdown quantum criticality in optical conductivity}
\date{\today}
\author{I. Paul$^{1,2}$, M. Civelli$^2$}
\affiliation{
$^1$ Institut N\'{e}el, CNRS/UJF, 25 avenue des Martyrs, BP 166, 38042 Grenoble, France\\
$^2$ Institut Laue-Langevin, 6 rue Jules Horowitz, BP 156, 38042 Grenoble, France
}

\begin{abstract}
We show that the behavior of the finite frequency inter-band transition
peak in the optical conductivity of the heavy fermions can provide definitive
experimental proof of the Kondo breakdown phenomenon in which the lattice Kondo
temperature vanishes at a quantum critical point. Approaching
such a phase transition from the heavy Fermi liquid side, we find a new cross-over
regime where the peak position is related to, but is not directly proportional to
the lattice Kondo scale. The peak position moves to lower energies but remains
finite, while the peak value changes non-monotonically and it eventually disappears
at the quantum critical point. These are unique signatures which distinguish a
Kondo breakdown transition from a spin density wave driven transition.

\end{abstract}

\maketitle

\emph{Introduction.}---
Developments in material synthesis have led to the
discovery of numerous rare earth compounds, the heavy fermions (HFs),
which can be tuned to a quantum critical point (QCP), separating a magnetic ground
state from a paramagnetic one~\cite{stewart},
by varying an external parameter such as pressure or chemical doping.
In the quantum critical regime the metallic properties are
significantly different from what one expects from a standard Landau
Fermi liquid (LFL), and therefore
these systems are prototypes to study how strong correlation effects
give rise to deviations from LFL.

The early theoretical attempts to describe the QCP are based on the
possibility that the instability of the paramagnetic phase is due to spin density
wave (SDW) formation~\cite{hmm}, and the critical fluctuations
are the paramagnons.
However, in three dimensions these theories fail to explain simultaneously
the linear temperature ($T$)
dependence of the resistivity and the $\log T$ dependence of the specific heat
coefficient observed in experiments~\cite{review-piers}.
This has stimulated theorists
to construct alternative descriptions of the quantum criticality~\cite{qimiao},
among which the scenario of
Kondo breakdown (KB) has proven to be promising~\cite{senthil,ppn}.
In the KB scenario
the non-LFL behaviour is due to the presence of a \emph{second} QCP, in
close proximity to the magnetic one, where the effective Kondo temperature of the
lattice ($T_K$) goes to zero.
In this picture the critical fluctuations, non-magnetic in origin, are associated with
the hybridization fluctuation between a broad conduction band and a narrow 
$f$-electron band.

In view of the current interest in
KB~\cite{senthil,ppn,schofield,cp-team},
it is apt to pose the question if 
a \emph{direct} experimental signature of the KB could distinguish it
from other possible routes to quantum criticality.
The issue is non-trivial partly due to the absence of low-$T$ angle-resolved
photoemission data of the HFs, and partly because standard quantities such as
resistivity and specific heat do not measure $T_K$ directly.
For example, the violation of the Wiedemann-Franz (WF) law and the divergence of the
Gr\"{u}neisen ratio (GR) at the QCP have been identified as experimental consequences of
KB~\cite{cp-team}. However, a true violation of the WF law signifies
the absence of electron quasiparticles which need not necessarily be due to
$T_K \rightarrow 0$ physics. An apparent violation of
the WF law near a SDW transition~\cite{mckenzie} can also take place because of
inelastic scattering.
Similarly, the divergence of the GR is expected for all kinds of QCPs~\cite{zhu}.
Consequently, while KB phenomenon implies the violation of the WF law and the
divergence of the GR, the converse does not hold.
The main purpose of this work is to point out that finite frequency features
in the optical conductivity of the HFs, the so-called mid-infrared peak
which arise due to inter-band optical transitions, can provide a direct and unique experimental
signature of the KB phenomenon.

The mid-infrared peak in the optical conductivity of the HFs
has been well-studied
experimentally~\cite{expt-mir} and theoretically~\cite{theory-mir}
in the case where the system is far from any phase instability. The
physical origin of the peak is readily understood within a two-band picture of the
Kondo effect in a lattice, which describes the formation of the heavy LFL as
the hybridization between a conduction $c$-band and a narrow $f$-band.
The inter-band transition takes place between one band with more
$f$-character and another with more $c$-character, and therefore it involves an
energy-scale \emph{at least} of the order of the binding energy $T_K$ of the composite
heavy electrons. Thus the peak position is related to $T_K$,
even though the two quantities may not always be proportional,
as we show below.

A crucial aspect of the SDW approach is that
the paramagnetic energy scale $T_K$
remains finite at the QCP.
Therefore, for such a transition one does not expect the
mid-infrared peak position and the peak height to change significantly while the system is
tuned to the QCP by varying
the external parameter.
In contrast, the KB phenomenon is based upon the possibility 
of a QCP where $T_K \rightarrow 0$.
Consequently, a hallmark of KB is a significant shift of the inter-band
peak position to lower energies, as the system is tuned to the QCP. Furthermore, since at the
KB-QCP the $f$-band effectively decouples from the $c$-band, one expects the inter-band
feature to gradually disappear.
These qualitative differences can be useful to distinguish experimentally a KB transition
from a SDW instability. This motivates us
to study how the mid-infrared peak for a heavy LFL varies as the system approaches a KB-QCP.

From the perspective of the charge-dynamics of the $f$-electrons,
the KB-QCP has been identified as an
orbital selective Mott transition,
where the $f$-electrons localize and decouple from the metallic
environment\cite{pepin-selective,lorenzo}.
Orbital selective localization is a relevant phenomenon in materials with
both weakly and strongly correlated bands, 
including classical Mott insulators (like VO$_{2}$~\cite{goodenough71}
and V$_{2}$O$_{3}$~\cite{ezhov99}), ruthenates Ca$_{2-x}$Sr$_{x}$RuO$_{4}$~\cite{anisimov02},
cobaltates, fullerenes and layered organic superconductors
(for a complete list see e.g. Ref.~\cite{imada98-kotliar06}).
KB is likely only one
of the several possible microscopic mechanisms that leads to orbital selective Mott transition. However it
is interesting to conjecture that, besides the HFs, there are other cases where
the localization can be described in terms of a vanishing ``effective
Kondo temperature''.
We expect that in such systems the study of the
optical conductivity will be equally germane. 

\emph{Model.}---
The system we study is described by the Anderson-Heisenberg model
which is given by
\bea
\label{eq:hamiltonian}
\ham
& = &
- \sum_{\langle ij \rangle, \si} \left(
t_c \cda_{i \si} c_{j \si} + t_f \fda_{i \si} f_{j \si}
\right)
- E_0 \sum_{i, \si} n_{i \si}^f
\nonumber \\
& + &
V_0 \sum_{i, \si} \left(  \cda_{i \si} f_{i \si} + {\rm h.c.} \right)
+ U \sum_i n_{i \uparrow}^f n_{i \downarrow}^f
\nonumber \\
& + &
J_H \sum_{\langle ij \rangle} \vec{S}_{i}^f \cdot \vec{S}_{j}^f.
\eea
Here $t_c$ and $t_f$ ($t_c \gg t_f$) denote hopping of $c$- and $f$-
electrons respectively,
($i$, $j$) are lattice sites (which are, say, nearest and next-nearest neighbours),
$E_0$ is the local energy level of the $f$-band,
and $\si$ is the spin index.
$V_0$ is the hybridization between the bands, $U$ is the on-site
Coulomb repulsion in the $f$-band
($U \gg t_f$),
and $J_H \sim (t_f)^2/U$ is a superexchange spin-spin interaction in the $f$-band.
We study the above
model in the Kondo limit where the $f$-electrons form local moments.

The mean field theory is generated by replacing
$f_{i \si} \rightarrow \bda_i F_{i \si}$,
where ($\bda_i$, $b_i$) describe charged holons and
($\Fda_{i \si}$, $F_{i \si}$) describe
spin-1/2 spinons. The large Coulomb repulsion is taken into account
by projecting out the double-occupied $f$-electron states via
the constraint $\sum_{\si} \Fda_{i \si} F_{i \si} + \bda_i b_i = 1$ 
(the associated Lagrange multiplier renormalizes $E_0$).
At the mean field level this model shows a KB
quantum phase transition at a critical value $x_c$ of the parameter
$x \equiv V_0/J_H$~\cite{pepin-selective}.
For $x > x_c$ one obtains a heavy LFL phase
where $\langle b_i \rangle \neq 0$ and the $f$-electrons participate in
band formation, and for $x < x_c$ one obtains a phase with $\langle b_i \rangle = 0$
where the $f$-electrons localize and form a non-magnetic uniform spin liquid phase (with
$\langle \Fda_{i \si} F_{j \si} \rangle \neq 0$). As the QCP is approached
from the LFL side, the lattice Kondo temperature $T_K \approx \pi \langle b \rangle^2 V_0^2 \nu_0
\rightarrow 0$, where $\nu_0$ is the bare density of states of the $c$-band.
We write the spinon bandwidth as $\al t_c \sim (\langle b \rangle^2 t_f + J_H)$,
and consider $\al \ll 1$ as a small parameter of the theory.

We use linear response theory to study the $T = 0$ optical conductivity
of the system as it approaches the QCP from the LFL side. 
We ignore the $f$-band contribution to the current operator and
write it as
$\hat{J}_{\mu} \approx \sum_{\bk \si} (\nabla_{\bk} \ep_{\bk})_{\mu} \cda_{\bk \sigma} c_{\bk \sigma}$,
where $\ep_{\bk}$ is the bare dispersion of the $c$-band, and $\mu$ denotes
spatial direction.
This does not change
our results at a qualitative level, and can be formally justified as an expansion in $t_c/t_f$.
Furthermore, we write $\left| \nabla_{\bk} \ep_{\bk} \right| \approx v_F$, the Fermi velocity
of the $c$-band, and we ignore the vertex corrections to the current-current correlator.
The optical conductivity
$\si_1 (\Om)$ is the real part of the frequency dependent conductivity tensor, which
can be written as (for $\Om \geq 0$, and setting $\hbar = 1$)
\ben
\label{eq:optical1}
\si_1(\Om) \! \approx \! \frac{2 e^2 v_F^2}{\pi \Om d} \! \int_{- \Om}^0 \! d \om
\! \sum_{\bk} {\rm Im} G_c^R(\om, \bk) {\rm Im} G_c^R(\om + \Om, \bk),
\een
where $R$ denotes the retarded function.
Within the mean field theory the Green's function for the $c$-electrons
is 
$ G_c^R (\om, \bk)^{-1} = \om - \ep_{\bk} + i/(2 \tau_c)
- \Sigma_c^R (\om, \bk)$,
$\Sigma_c^R (\om, \bk) = V^2/[\om - \ep^0_{\bk} + i/(2 \tau_F) ]$.
$V = \langle b \rangle V_0$ is the effective hybridization between the $c$-band
and the $F$-band of the spinons, $\ep^0_{\bk}$ is the dispersion of the spinons whose
bandwidth is $\al t_c$, and 
$\tau_c$ and $\tau_F$ mimic the elastic scattering due to the presence of impurities. 
In principle, $\al t_c$ is $V$ dependent, but, for simplicity, we treat it as a constant.

\begin{figure}[!!t]
\begin{center}
\includegraphics[width=9cm,height=4.50cm,angle=-0]{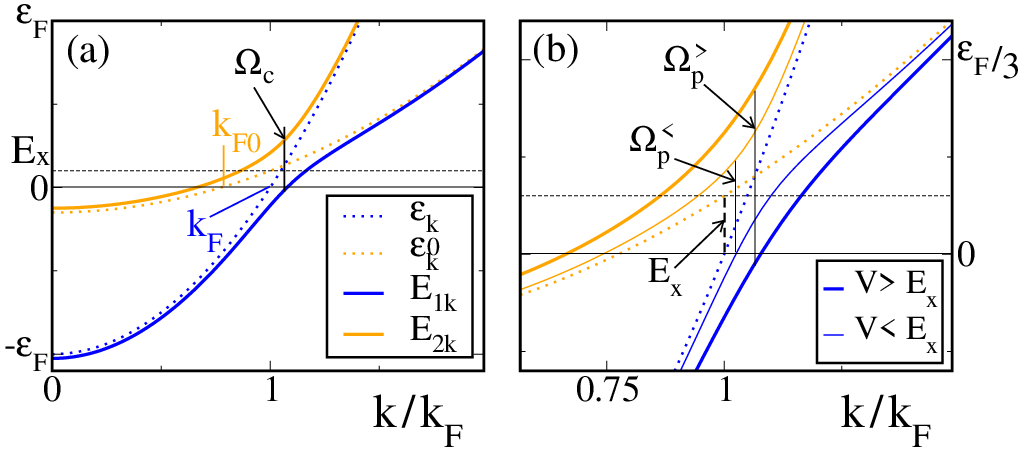}
\caption{(color online). (a) $\ep_{\bk}$ and $\ep^0_{\bk}$,
are the bare conduction and spinon bands
(with respective Fermi vectors $k_F$ and $k_{F0}$).
$E_x= \ep^0_{k_F}$ is a microscopic energy-scale.
$E_{1,2 \bk}$ are the hybridized bands, $V$ is the effective hybridization
and $\Om_c = 2V$ is the minimal inter-band transition-energy
(corresponds to momenta for which $\ep_{\bk} = \ep^0_{\bk}$).
(b) For $V > E_x$, the optical conductivity peak is at $\Om_p^> = \Om_c$.
For $V < E_x$ the $\Om_c$-transition is forbidden at $T=0$,
(since $E_{1 \bk}> 0$ is above the chemical potential).
The peak is at $\Om_p^< = V^2/E_x + E_x$.}
\label{fig1}
\end{center}
\end{figure}

\begin{figure}[!!!t]
\begin{center}
\includegraphics[width=8cm,height=6.0cm,angle=-0]{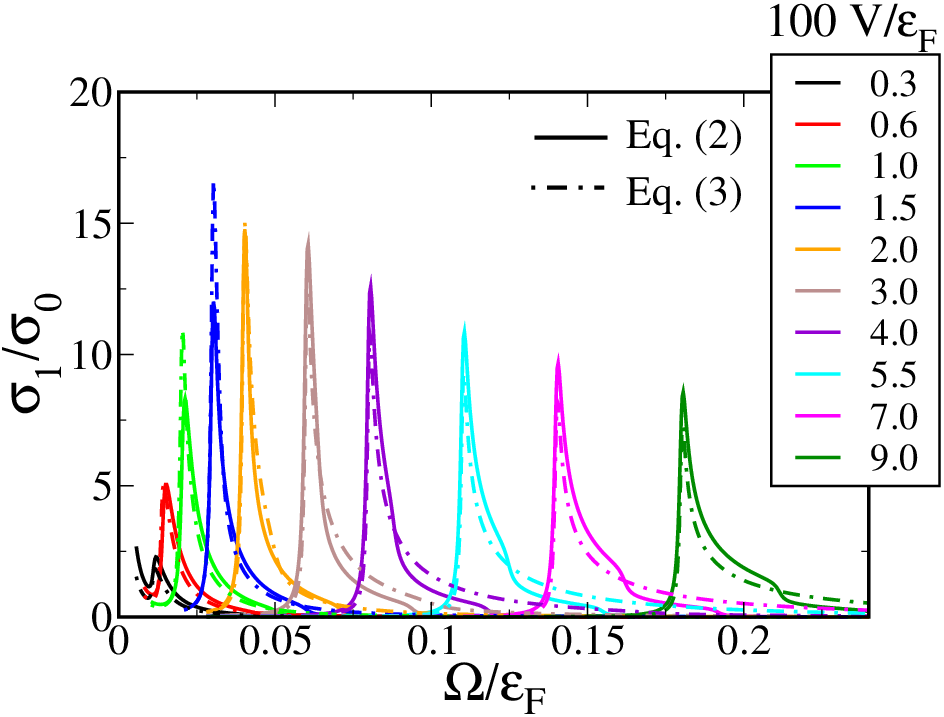}
\caption{(color online).
Evolution of the optical conductivity $\si_1(\Omega)$ with effective hybridization $V$
ranging from $0.3 \times 10^{-2}\varepsilon_{F}$ (left) to
$9 \times 10^{-2}\varepsilon_{F}$ (right).
$\si_1(\Omega)$ is expressed in unit of $\si_0 = e^2/(m \ep_F)$.
The solid lines (Eq.~\ref{eq:optical1}) are obtained using parabolic
bands ($\alpha=0.25$ and $1/\tau_{c[F]}=10^{-3} \ep_{F}$).
The dot-dash lines (Eq.~\ref{eq:optical2}) are obtained using
a linearized $c$-band and a non-dispersive $F$-band.
In both cases, as $V$ is reduced, the peak heights
first increase for $V \gg E_x=\ep_{F}/100$, and then diminish for $V \lesssim E_x$.
For clarity the plots are truncated to exclude the Drude feature.}
\label{fig2}
\end{center}
\end{figure}

\begin{figure}[!!!t]
\begin{center}
\includegraphics[width=8.50cm,height=4.0cm,angle=-0]{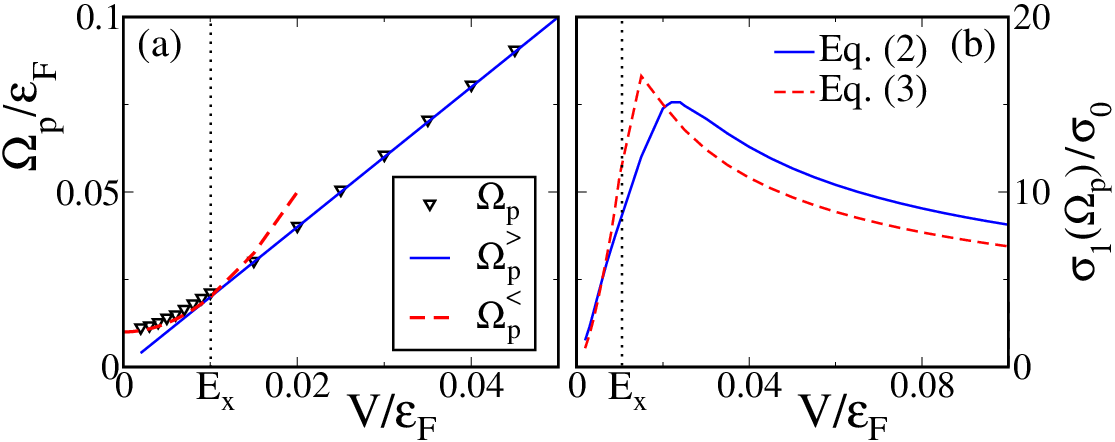}
\caption{(color online). (a) The position $\Om_p$ of the inter-band peak
in $\si_1 (\Om)$, and (b) the peak height $\si_1(\Om_p)$ as a function of
the effective hybridization $V$.
The triangles in (a) and the solid line in (b)
are obtained using parabolic bands (Eq.~\ref{eq:optical1} with $\alpha=0.25$
 and  $1/\tau_{c[F]}=10^{-3} \ep_F$).
The peak position matches the analytical result $\Om_p = \Om_p^>$ ($\Om_p^<$)
for $V > E_x$ ($V < E_x$) . In (b) the dashed line is obtained
using a linearized $c$-band and non-dispersive $F$-band (Eq.~\ref{eq:optical2}).
}
\label{fig3}
\end{center}
\end{figure}

We adopt few simplifications
in order to perform the calculations analytically:
(i) the $c$-band dispersion is linearized $\ep_{\bk} = v_F (k - k_F)$,
where $k_F$ is the Fermi wave-vector of the $c$-band;
(ii) the mildly dispersive $F$-band is replaced by a flat level
$\ep^0_{\bk} \approx \ep^0_{k_F} \equiv E_x$ [Fig. \ref{fig1}(a)].
Here $E_x  \approx \al v_F (k_F - k_{F0})$, where $k_{F0}$
is the Fermi wave-vector of the $F$-band, and $k_{F0} < k_F$. 
The latter approximation is equivalent to ignoring all non-singular $\al$
dependence in Eq.~(\ref{eq:optical1}).
The microscopic non-universal parameter $E_x$, which is the energy
necessary for a momentum conserving inter-band transition from the Fermi surface
of the $c$-band to the $F$-band, becomes an important energy-scale
in the limit where the effective hybridization $V$ is
small (see below). 
We do not expect our results to depend
on the 
details of the band-dispersions. To demonstrate this explicitly,
we also evaluate numerically $\sigma_{1}(\Omega)$ in Eq. \ref{eq:optical1} by taking
the continuum limit of the model.
We use parabolic dispersions
$\ep_{\bk} = (k^2 - k_F^2)/(2m)$ and $\ep^0_{\bk} = (k^2 - k_{F0}^2)/(2m_0)$,
set the energy scale by $\ep_{F} \equiv k_F^2/(2m)=1$, and we choose $E_x= \ep_{F}/100$,
$m/m_0 \equiv \al= 0.25$
We show that
at a qualitative level the numerical and approximate analytical results
match well
[Figs. \ref{fig2}, \ref{fig3}], and this validates our physical conclusions.

\emph{Results.}---
Let us for the moment ignore the finite
lifetimes of the fermions and write
\[
{\rm Im} G_c^R(\om, \bk) = - \pi [ u_{\bk}^2 \dl (\om - E_{1 \bk}) + v_{\bk}^2 \dl (\om - E_{2 \bk})],
\]
where $u_{\bk}^2 = [1 + (E_x - \ep_{\bk})/D]/2$,
$D = [(\ep_{\bk} - E_x)^2 + 4 V^2]^{1/2}$,
$v_{\bk}^2 = 1- u_{\bk}^2$, and
$E_{1,2 k} = [\ep_{\bk} + E_x \mp D]/2$ are the two hybridized bands. From
Eq.~(\ref{eq:optical1}) we get (for $\Om > 0$),
\[
\si_1(\Om) \propto \frac{1}{\Om} \int_{- \Om}^0 d \om \sum_{\bk}
u_{\bk}^2 v_{\bk}^2 \dl (\om - E_{1 \bk}) \dl (\om + \Om - E_{2 \bk}),
\]
which is finite provided (a) $- \Om \leq E_{1 \bk} \leq 0$, and
(b) $\Om = E_{2 \bk} - E_{1 \bk}$. Thus, any finite frequency feature in the
optical conductivity is due to momentum conserving inter-band transitions.
We note that, for a given effective hybridization
$V$, there is a cut-off frequency
$\Om_c \equiv \left[ E_{2k} - E_{1k} \right]_{\rm min} = 2 V$
below which inter-band transitions are not possible. This corresponds to
momenta where 
the bare bands
$\ep_{\bk}$ and $\ep^{0}_{\bk}$
cross [Fig. \ref{fig1}(a)]. For this process 
$u_{\bk}^2 v_{\bk}^2 \approx 1/4$ is maximal, and it reduces
for inter-band transitions with frequencies larger than $\Om_c$.

In order to include effects of finite lifetime of the
$c$-electrons (we find that finite $\tau_F$ is of secondary importance, and we ignore
it in the analytic evaluation),
it is more convenient to replace the momentum sum in Eq.~(\ref{eq:optical1})
by an energy integral, which can be performed by the method of
contours for the simplified case of a linearized $c$-band and a
non-dispersive $F$-band. This gives
\ben
\label{eq:optical2}
\si_1(\Om) \approx - \frac{\Om_0^2}{\Om} {\rm Im} \int_{- \Om}^{0}
\frac{d \om}{\Om + i/\tau_c+ \Sigma^A_c(\om) - \Sigma^R_c(\om + \Om)},
\een
where $\Om_0^2 = ne^2/m$, $n$ being the  $c$-electron density.
As $V$ is reduced we find 
two different regimes (Figs. \ref{fig2} and \ref{fig3}).

(i) $V > E_x$. In this regime $\Om =  \Om_c$ corresponds to $E_{1 \bk} = E_x-  V < 0$,
which implies that the threshold frequency for an inter-band transition is 
$\Om_c$
[Fig. \ref{fig1}(a)]. Close to the threshold we get
(dot-dashed line in Fig. \ref{fig2})
\bea
\si_1(\Om \gtrsim \Om_c)
&\approx&
\frac{\Om_0^2 \Om^2}{4 \Om_c^2} {\rm Im} \left[ \frac{1}{2Z}
\left\{ \ln \left(1- \frac{\Om}{\Om/2 - E_x + Z}\right)
\right. \right.
\nonumber \\
&-&
\left. \left.
\ln \left(1- \frac{\Om}{\Om/2 - E_x - Z}\right)
\right\}
\right],
\nonumber
\eea
where $Z = (\Om_+^2 + i/\tau_1^2)^{1/2}$, $\Om_+ = \Om (\Om^2 - \Om_c^2)^{1/2}/(2 \Om_c)$,
and $1/\tau_1^2 = \Om^3/(4 \tau_c \Om_c^2)$.
The difference in the position of the peak $\Om_p$ in $\si_1 (\Om)$ and the
threshold $\Om_c$ is $\ord(1/(\Om_c \tau_c))$,
and assuming well defined quasiparticles with $\Om_c \tau_c \gg 1$ we get
[  Fig. \ref{fig1}(b) and \ref{fig3}(a)]
\ben
\Om_p \approx \Om_p^> \equiv 2V.
\een
Thus, in this regime, which has been discussed in the literature~\cite{expt-mir,theory-mir},
$\Om_p \propto \sqrt{T_K}$ and the peak position is a direct measure of the lattice Kondo
temperature.
For $V \gg E_x$, 
$\si_1 (\Om_p) \propto \Om_0^2 [\tau_c/V]^{1/2}$, which implies
that the peak value \emph{increases} as $V$ is reduced
[Fig. \ref{fig2} and Fig. \ref{fig3}(b)].
This behaviour continues until $(\tau_c/V)^{1/2} (V-E_x) \sim 1$,
after which the peak value \emph{decreases} with decreasing $V$.
We note that, the divergence of $\si_1(\Om_p)$ for infinitely long-lived $c$-electrons is
an artefact of a theory where the
effective hybridization is momentum independent. For the case where the momentum dependence of the
hybridization is important, we expect $\si_1(\Om_p) \propto 1/V$.

(ii) $V < E_x$. This new regime, which has not been studied earlier, is relevant when
the system is close to the KB-QCP.
We find that $\Om =  \Om_c$ corresponds to $E_{1 \bk} = E_x - V > 0$,
which implies that an inter-band transition at this frequency is not possible at $T=0$.
The threshold frequency is $\Om_p^< \equiv V^2/E_x + E_x > \Om_c$,
and this corresponds to an inter-band transition from the Fermi surface of the
lower hybridized band [Fig. \ref{fig1}(b)].
In the vicinity of the threshold we get
\[
\si_1 (\Om \gtrsim V^2/E_x + E_x) \approx
\frac{\pi \Om_c^2}{4 \Om^2 \sqrt{\Om^2 - \Om_c^2}}.
\]
As before, neglecting the finite lifetime, 
the peak-position is at 
the threshold frequency,
and we get [Fig. \ref{fig3}(a)]
\ben
\Om_p \approx \Om_p^< = V^2/E_x + E_x.
\een
Thus, $\Omega_p$ is no longer a direct measure of $T_K$,
and in particular it stays finite ($\Om_p \rightarrow E_x$ as $T_K \rightarrow 0$)
at the QCP. Also, in this regime $\Omega_p$ varies little with $V$, which may not be
discernible whenever 
$\Omega_p$ cannot be very well-resolved.
Since $\Om_p \neq \Om_c$, the corresponding matrix element
$u_{\bk}^2 v_{\bk}^2$ decreases from the maximal value 1/4 as $V$ is reduced.
As a result the peak-value $\sigma_1(\Omega_p)$ \emph{decreases} rapidly
[Fig. \ref{fig2} and Fig. \ref{fig3}(b)], and the peak vanishes at the QCP
indicating the gradual decoupling of the $f$-band from the $c$-band. 
Note that if $1/\tau_c \gg E_x$, the inter-band feature is eventually
masked by the Drude peak.

Next we discuss the limitations of our results. These are based on
model calculations and mean field considerations giving more
prominent inter-band feature than those observed experimentally.
Furthermore, the fluctuations are non-negligible near the QCP, and
the crucial question is whether this only smears the feature further
or washes it out entirely. The issue requires further investigation
that is beyond the scope of the current paper. However, even if the
feature is washed out very near the QCP, any observation
of significant shift of the peak position towards lower energy will
have demonstrated that the physics of these systems is richer than what
can be captured by SDW theory.

\emph{Conclusion.}---
We studied the finite-frequency inter-band transition peak
in the optical conductivity of
a heavy fermion system near a Kondo breakdown type of quantum critical point,
where the lattice Kondo temperature vanishes.
As the system approaches the QCP from the heavy Fermi liquid
side, we find a new cross-over regime where the peak position is related to, but is no longer a direct
measure of the lattice Kondo scale. In particular,
the peak position moves to lower energies,
but remains finite at the QCP. On the other hand, the peak value changes
non-monotonically as $T_K \rightarrow 0$, and eventually the peak disappears
at the QCP indicating the decoupling of the $f$-electrons
from the conduction band. These features are fingerprints
of a Kondo breakdown type of QCP, and therefore can be used to
distinguish it experimentally 
from a spin density wave type of QCP.


\begin{thebibliography}{99}

\bibitem{stewart}
G. Stewart, Rev. Mod. Phys. {\bf 73}, 797 (2001);
H. von L\"{o}hneysen, A. Rosch, M. Vojta and P. W\"{o}lfle, Rev. Mod. Phys. {\bf 79}, 1015 (2007).

\bibitem{hmm}
J. A. Hertz, Phys. Rev. B \textbf{14}, 1165 (1976);
T. Moriya, \textit{Spin Fluctuations in Itinerant Electron
Magnetism}, (Springer-Verlag, Berlin, New York, 1985);
A. J. Millis, Phys. Rev. B \textbf{48}, 7183 (1993).

\bibitem{review-piers}
see e.g.,
P. Coleman, C. Pepin, Q. Si and R. Ramazashvili, J. Phys.: Condens. Matter {\bf 13}, R723 (2001).

\bibitem{qimiao}
Q. Si, S. Rabello, K. Ingersent and J. L. Smith, Nature (London) {\bf 413}, 804 (2001).

\bibitem{senthil}
T. Senthil, S. Sachdev, and M. Vojta, Phys. Rev. Lett. {\bf 90}, 216403 (2003);
T. Senthil, M. Vojta, and S. Sachdev, Phys. Rev. B {\bf 69}, 035111 (2004).

\bibitem{ppn}
I. Paul, C. P\'{e}pin and M. R. Norman, Phys. Rev. Lett. {\bf 98}, 026402 (2007);
Phys. Rev. B 78, 035109 (2008).

\bibitem{schofield}
P. Coleman, J. B. Marston and A. J. Schofield, Phys. Rev. B {\bf 72}, 245111 (2005).

\bibitem{cp-team}
A. Benlagra, and C. P\'{e}pin, Phys. Rev. Lett. {\bf 100}, 176401 (2008);
K.-S. Kim, A. Benlagra, and C. P\'{e}pin, Phys. Rev. Lett. {\bf 101}, 246403 (2008);
K.-S. Kim, and C. P\'{e}pin, Phys. Rev. Lett. {\bf 102}, 156404 (2009).

\bibitem{mckenzie}
M. F. Smith, and R. H. McKenzie, Phys. Rev. Lett. {\bf 101}, 266403 (2008).

\bibitem{zhu}
L. Zhu, M. Garst, A. Rosch, and Q. Si,
Phys. Rev. Lett. {\bf 91}, 066404 (2003).

\bibitem{expt-mir}
see e.g., S. R. Garner, J. N. Hancock, Y. W. Rodriguez, Z. Schlesinger,
B. Bucher, Z. Fisk, and J. L. Sarrao, Phys. Rev. B {\bf 62}, R4778 (2000);
S. V. Dordevic, D. N. Basov, N. R. Dilley, E. D. Bauer, and M. B. Maple,
Phys. Rev. Lett. {\bf 86}, 684 (2001);
F. P. Mena, D. van der Marel, and J. L. Sarrao,
Phys. Rev. B {\bf 72}, 045119 (2005);
H. Okamura, T. Watanabe, M. Matsunami, T. Nishihara, N. Tsujii, T. Ebihara,
H. Sugawara, H. Sato, Y. Onuki, Y. Isikawa, T. Takabatake, and T. Nanba,
J. Phys. Soc. Jpn. {\bf 76}, 023703 (2007).

\bibitem{theory-mir}
P. Coleman, Phys. Rev. Lett. {\bf 59}, 1026 (1987);
L. Degiorgi, F. B. B. Anders, and G. Gr\"{u}ner, Eur. Phys. J. B {\bf 19}, 167 (2001);
J.H. Shim, K. Haule, and G. Kotliar, Science {\bf 318}, 1615 (2007);
H. Weber, and M. Vojta, Phys. Rev. B {\bf 77}, 125118 (2008).

\bibitem{pepin-selective}
C. P\'{e}pin, Phys. Rev. Lett. {\bf 98}, 206401 (2007);
Phys. Rev. B {\bf 77}, 245129 (2008).

\bibitem{lorenzo}
L. De Leo, M. Civelli, and G. Kotliar,
Phys. Rev. B {\bf 77}, 075107 (2008); Phys. Rev. Lett. {\bf 101}, 256404 (2008).

\bibitem{goodenough71}
J.B. Goodenough, Prog. Solid State Chem. {\bf 5}, 145 (1971).

\bibitem{ezhov99}
S.Y. Ezhov, V.I. Anisimov, D.I. Khomskii, and G.A. Sawatzky,
Phys. Rev. Lett. {\bf 83}, 4136 (1999).

\bibitem{anisimov02}
V.I. Anisimov, I.A. Nekrasov, D.E. Kondakov, T.M. Rice, and M. Sigrist,
Eur. Phys. J. B {\bf 25}, 191 (2002).

\bibitem{imada98-kotliar06}
Imada, M., A. Fujimori, and Y. Tokura, 1998, Rev. Mod.
Phys. {\bf 70}, 1039; see e.g. sec. III.3 in
G. Kotliar, S. Y. Savrasov, K. Haule, V. S. Oudovenko, O. Parcollet, and C. A. Marianetti,
Rev. Mod. Phys. {\bf 78}, 865 (2006).

\end{thebibliography}
\end{document}